\documentclass[aps, twocolumn,prb,showpacs,preprintnumbers,amsmath,amssymb,superscriptaddress]{revtex4-1}
\usepackage{amsmath,amsfonts,amssymb,graphics,graphicx,epsfig,color,times,indentfirst,layout,hyperref,subfigure,multirow}
\usepackage{hyperref}
\usepackage{ulem}
\hypersetup{linktocpage}
\usepackage{threeparttable}

\usepackage{titlesec}

\begin{document}

\title{
Majorana vortex-bound states in three-dimensional nodal noncentrosymmetric superconductors
 }

\author{Po-Yao Chang}
\affiliation{Department of Physics, University of Illinois at Urbana-Champaign, Urbana, Illinois 61801, USA}

\author{Shunji Matsuura}
\affiliation{Department of Physics and Mathematics, McGill University, 
Montr\'eal, Qu\'ebec, Canada}
\author{Andreas P. Schnyder}

\affiliation{Max-Planck-Institut f\"ur Festk\"orperforschung,  
Heisenbergstrasse 1, D-70569 Stuttgart, Germany}

\author{Shinsei Ryu}
\affiliation{Department of Physics, University of Illinois at Urbana-Champaign, Urbana, Illinois 61801, USA}

\date{\today}

\begin{abstract}
Noncentrosymmetric superconductors (NCSs), characterized by  antisymmetric spin-orbit coupling and a mixture of spin-singlet and spin-triplet pairing components, are promising candidate materials for topological superconductivity. 
An important hallmark of topological superconductors is the existence of protected zero-energy states at surfaces or in vortex cores.
Here we investigate Majorana vortex-bound states in three-dimensional nodal and fully gapped NCSs 
by combining analytical solutions of Bogoliubov-de Gennes  (BdG) equations in the continuum with exact diagonalization of BdG Hamiltonians.
We show that depending on the crystal point-group symmetries and the topological properties of the  
bulk Bogoliubov-quasiparticle wave functions, different types of zero-energy Majorana modes can appear inside the vortex core.
We find that for nodal NCSs with tetragonal point group $C_{4v}$ the vortex states are dispersionless along the vortex line,
forming one-dimensional Majorana flat bands, while for NCSs with $D_{4}$ point-group symmetry 
the vortex modes are helical Majorana states with a linear dispersion along the vortex line.
NCSs with monoclinic point group $C_2$, on the other hand, do not exhibit any zero-energy vortex-bound states.
We show that in the case of the $C_{4v}$ ($D_4$) point group the stability of these Majorana zero modes is guaranteed by a combination of reflection ($\pi$ rotation), time-reversal, and particle-hole symmetry. 
Considering continuous deformations of the quasiparticle spectrum in the presence of vortices, we show
that the flat-band vortex-bound states of $C_{4v}$ point-group NCSs can be adiabatically connected to the dispersionless vortex-bound states 
of time-reversal symmetric Weyl superconductors. 
Experimental implications of our results for thermal transport and tunneling measurements are discussed.
\end{abstract}

\maketitle

\noindent

\section{Introduction}
Topological superconductors have in recent years become a subject of intense research
due to their potential use for technical applications in device fabrication and quantum information \cite{Hasan, Ryu2010ten, Qi,beenakkerReview11,aliceaReview12}.
By the bulk-boundary correspondence, zero-energy Majorana modes appear at the surface or inside
the vortex core of topological superconductors.
The experimental search for  Majorana modes, which can be utilized as basic building blocks (i.e., half a qubit) for topological quantum computers,
is the focus of a growing research effort~\cite{Mourik, Das, Deng, Eduardo}.
These Majorana modes are robust against symmetry preserving impurity scattering processes and deformations of the superconducting order parameter.
While topological superconductivity can be artificially engineered in heterostructures with an ordinary $s$-wave superconductor and, say, a semiconductor~\cite{sau,lutchyn,oreg} or a topological insulator~\cite{fuKanePRL08}, it can also occur naturally  in certain correlated materials with strong spin-orbit coupling (SOC).

One promising class of materials for topological superconductivity are the noncentrosymmetric superconductors (NCSs) \cite{bauerSigristBook}.
In these systems, the absence of inversion symmetry together with strong SOC and electronic correlations can give rise to
 unconventional pairing states with topologically nontrivial characteristics \cite{SatoPRB06,Sato3,Tanaka,Sato2,Beri,Schnyder,Yada,Brydon,Matsuura,Schnyder1}. For example, 
 in CePt$_3$Si~\cite{bauer2004heavy,Izawa,Bonalde}, macroscopic as well as microscopic measurements indicate an unconventional superconducting state
 with a mixture of spin-singlet and spin-triplet pairing components and line nodes in the superconducting gap.
 Experimental evidence for unconventional pairing symmetries has also been reported for  CeIrSi$_3$ \cite{Mukuda}, CeRhSi$_3$ \cite{kimuraPRL05}, Y$_2$C$_3$~\cite{chenPRB11}, Li$_2$Pt$_3$B \cite{Yuan, Nishiyama, Eguchi}, and BiPd \cite{Mondal}.
Both fully gapped and nodal NCSs with sizable spin-triplet pairing components exhibit nontrivial topological properties,
which manifest themselves in terms of different types of zero-energy  surface states.
In fully gapped NCSs the surface states are dispersing helical Majorana modes, whereas nodal NCSs exhibit flat-band surface states~\cite{Schnyder,Brydon,Sato1,Tanaka,Yada,Matsuura}, 
and depending on the crystallographic point group, may also support helical Majorana modes or arc surface states~\cite{Schnyder1}.
Experimentally, it is possible to distinguish among different types of surface states using
Fourier-transform scanning tunneling spectroscopy \cite{HofmannPRB13} or 
 surface transport measurements \cite{Schnyder2,brydonNJP13}.
 
Most of the candidate materials for noncentrosymmetric topological superconductivity are strong type-II superconductors, with 
Ginzburg-Landau parameters $\kappa$ of the order of $\sim$100~\cite{bauerSigristBook}.
Hence, zero-energy Majorana modes may emerge inside magnetic vortices of these superconducting compounds~\cite{Lu,Fuji08,Kashyap,Sato3}.
In this paper, we examine vortex-bound states of three-dimensional (3D) NCSs and 
study how their appearance is related to the crystal point-group symmetries of the superconductor and the nontrivial topological properties 
of the bulk Bogoliubov-quasiparticle wave functions.
Using both numerical and analytical methods, we compute the vortex-bound state spectra of $(s+p)$-wave NCSs 
with three different point-group symmetries: the two tetragonal point-groups $D_4$ and $C_{4v}$,
as well as the monoclinic point-group $C_2$~\cite{footnote_vortex}. 

One of our primary findings is that $D_4$ point-group NCSs support gapless
helical Majorana states inside vortex cores.
These subgap states disperse linearly along the vortex line,  
and are akin to one-dimensional helical Majorana modes that exist 
at the edge of  fully gapped topological NCSs in two dimensions. 
Remarkably, these vortex-bound states appear both in the fully gapped topological phase 
and in the nodal phase that separates the fully gapped trivial phase from the topological one [see Figs.~\ref{fig:phase_diagram}(a) and~\ref{fig: Scenario AI}].
While these helical Majorana vortex states exist in an extended 
region of the phase diagram of Fig.~\ref{fig:phase_diagram}(a), 
they are unstable against perturbations that break the $D_4$ point-group symmetry of the superconductor (SC) \cite{footnote_T_breaking}. 
For NCSs with tetragonal point-group symmetry $C_{4v}$, on the other hand, we find that
there are zero-energy vortex-bound states which are dispersionless along the vortex line,
forming a one-dimensional Majorana flat band (Fig.~\ref{fig: Scenario AII}).
In contrast, $C_2$ point-group NCSs do not exhibit any zero-energy vortex-bound states
neither in the fully gapped nor in the nodal phase (Fig.~\ref{fig: Scenario B}).
We find that the Majorana vortex-bound states of the $C_{4v}$ ($D_4$) point-group NCS
are protected by a combination of reflection ($\pi$ rotation), time-reversal, and particle-hole symmetry.

Interestingly, the existence of these vortex-bound states in nodal NCSs correlates 
to some degree 
with the appearance of extra surface states, that appear in addition to  the flat-band surface states.
That is, for nodal NCSs with $D_4$ point-group symmetry the helical vortex-bound states
always appear together with  helical Majorana cones on the surface which are
protected by a $\mathbb{Z}_2$ topological number \cite{Schnyder,Schnyder1}.
(In the following, we refer to these Majorana cone surface states as the ``$\mathbb{Z}_2$ surface states.")
On the other hand, for nodal $C_{4v}$ point-group NCSs the existence of flat-band vortex states
is correlated with  the appearance of helical arc states on the surface \cite{Schnyder1}, see  Table~\ref{tab:1}.
These  arc surface states are  superconducting analogues of the Fermi arcs that exist on the surface of Weyl semimetals~\cite{Wan, Burkov, Xu}.
Using translation symmetry in the vortex direction, we fix the momentum along the vortex line and
consider adiabatic deformations of the quasiparticle spectrum that 
do not close the bulk energy gap for this fixed momentum. 
By use of this procedure,
we find that the vortex-bound states  (extra surface states) of $D_4$  and $C_{4v}$ point-group NCSs
are adiabatically connected to the vortex-bound states (surface states) of fully gapped topological SCs and 
time-reversal symmetric Weyl SCs, respectively. Conversely, 
finite-energy vortex-bound states of nodal NCSs with $C_2$ point-group symmetry can be related 
to finite-energy vortex-bound states of fully gapped  trivial SCs (cf.\ Table~\ref{tab:1}).

\section{Model Hamiltonian and symmetries}
To study the appearance of vortex-bound states in nodal NCSs, we 
consider a generic single-band Bogoliubov-de Gennes (BdG)
Hamiltonian 
$H=\sum_{{\bf k}\in \mathrm{BZ}}    \Psi_{\bf k}^\dagger \mathcal{H}({\bf k} )   \Psi^{\phantom{\dag}}_{\bf k}$, 
with
\begin{align}
\mathcal{H}( {\bf k} )=&
\left(\begin{array}{cc}h ( {\bf k}) & \Delta( {\bf k}) \\ \Delta^\dagger( {\bf k})& -h^T (-{\bf k})\end{array}\right),
\label{Eq:BdG}
\end{align}
and the Nambu spinor
$\Psi_{\bf k}=( c^{\phantom{\dag}}_{{\bf k} \uparrow} , c^{\phantom{\dag}}_{{\bf k} \downarrow} , c^\dagger_{- {\bf k} \uparrow}, c^\dagger_{- {\bf k} \downarrow} )^{\mathrm{T}}$, where
$c^{\phantom{\dag}}_{{\bf k}\sigma}$ ($c^{\dag}_{{\bf k}\sigma}$) denotes
the electron annihilation (creation) operator
with momentum ${\bf k}$ 
and spin $\sigma=\uparrow,\downarrow$.
The normal-state Hamiltonian $h( {\bf k})=\varepsilon( {\bf k} )\mathbb{I}_{2\times 2}+\alpha {\bf l}( {\bf k} )\cdot \pmb{\sigma}$
describes electrons on a cubic lattice with nearest-neighbor hopping $t$, chemical potential $\mu$,
spin-independent dispersion
$\varepsilon( {\bf k} )=t(\cos k_x+\cos k_y+\cos k_z)-\mu$, 
and Rashba-type SOC $\alpha {\bf l}( {\bf k} )\cdot \pmb{\sigma}$ with strength $\alpha$. 
Here, $\pmb{\sigma}=(\sigma_1,\sigma_2,\sigma_3)$ is the vector of Pauli matrices.
Due to the absence of inversion symmetry, the superconducting gap $\Delta ( {\bf k} )$ contains in general an admixture of even-parity spin-singlet and odd-parity spin-triplet pairing components,
$\Delta({\bf k} )= (\Delta_s \mathbb{I}_{2\times 2} + \Delta_t  {\bf d}({\bf k} )\cdot \pmb{\sigma} ) (i \sigma_2)$,
where $\Delta_s$ and $\Delta_t$ represent the spin-singlet and spin-triplet pairing amplitudes, respectively. 
For the spin-triplet pairing term we assume that the vector ${\bf d}({\bf k} )$ is oriented parallel to the 
polarization vector ${\bf l}({\bf k})$ of the SOC \cite{frigeri04}.
To simplify matters we will set
 $( t, \alpha, \Delta_t) = (-1,1,1)$ 
 in our numerical calculations
 and study the
 vortex-bound states as a function of $\Delta_s$,  $\mu$, and
 different types of SOC potentials.
We have checked that different values of $( t, \alpha, \Delta_t)$  do not qualitatively change our results.
With   $\varepsilon( {\bf k} )=\varepsilon(-{\bf k})$ and ${\bf l}( {\bf k} )=- {\bf l}(- {\bf k} )$,
Hamiltonian~\eqref{Eq:BdG} is invariant under both time-reversal symmetry (TRS) and particle-hole symmetry (PHS), 
\begin{subequations}
\begin{eqnarray}
U_T^{-1}\mathcal{H}( {\bf k} )U_T =\mathcal{H}^*(-{\bf k} )  
\end{eqnarray}
and
\begin{eqnarray}
U_P^{-1}\mathcal{H}({\bf k} )U_P =-\mathcal{H}^*(-{\bf k} ),
\end{eqnarray}
\end{subequations}
where 
$U_T= \mathbb{I}_{2\times 2}\otimes i \sigma_2$ and $U_P=\sigma_1 \otimes \mathbb{I}_{2\times 2}$, respectively.
Hence, since $U^{\phantom{\ast}}_T U^{\ast}_T = - \mathbb{I}_{4\times 4}$ and $U^{\phantom{*}}_P U^{\ast}_P =  \mathbb{I}_{4\times 4}$, 
$\mathcal{H}( {\bf k} )$ belongs to symmetry class DIII.

\begin{table}
\begin{center}
\caption{ \label{tab:1}  
Depending on the crystal point-group symmetries (first column), nodal NCSs  can exhibit different types of zero-energy
vortex-bound states (second column). As indicated in the third column, the appearance of these different vortex states correlates with the existence
of extra surface states besides the flat-band states.
The helical vortex states and the $\mathbb{Z}_2$ surface states of  nodal $D_4$ NCSs can be adiabatically connected to the vortex-bound and surface states
of  fully gapped topological NCSs. Similarly,  the flat-band vortex states and arc surface states of $C_{4v}$ NCSs are related 
to the vortex-bound and surface states of  time-reversal symmetric Weyl SCs.
}
\begin{threeparttable}
\begin{tabular}{| c| c| c| c| }
\hline
           & Vortex states              & Extra surface states    & Adiabatic deformation  \\\hline
$D_4$      & helical states  & $\mathbb{Z}_2$ Majorana cone & fully gapped top.\ SC \\\hline
$C_{4v}$                 &  flat bands       &   helical arc states   & Weyl SC with TRS    \\\hline
$C_2$${}^{\textrm{a}}$                    & none            &   none    & gapped trivial SC \\\hline 
\end{tabular}
\begin{tablenotes}
\item[${}^{\textrm{a}}$]  for phase IV in Fig. 1(c).
\end{tablenotes}
\end{threeparttable}
\end{center}
\end{table}

The specific form of the spin-orbit coupling vector ${\bf l}(k)$ is constrained by the lattice symmetries  
of the superconductor~\cite{Samokhin}. In the following we consider NCSs with three different crystal point-group symmetries:
 the tetragonal point groups $D_4$ and $C_{4v}$, as well as  the monoclinic point group $C_2$.
Within a tight-binding expansion, we obtain for the crystal point group $D_4$ to lowest order
\begin{subequations}
\begin{equation}
{\bf l}( {\bf k} ) =(a_1\sin k_x, a_1\sin k_y, a_2\sin k_z).
\end{equation}
For the tetragonal point group $C_{4v}$,
which is relevant
for CePt$_3$Si, CeRhSi$_3$, and CeIrSi$_3$,  the vector ${\bf l}( {\bf k} )$ takes the form
\begin{equation} \label{SOCforC4v}
{\bf l}( {\bf k} ) = a_1 ( \sin k_y, -  \sin k_x, 0 ) .
\end{equation}
The lowest order terms compatible with $C_{2}$ point-group symmetry
(represented by BiPd) are given by
\begin{equation}
{\bf l}( {\bf k} ) =(a_1\sin k_x +a_5\sin k_y, a_2\sin k_y +a_4\sin k_x, a_3\sin k_z).
\end{equation}
\end{subequations}

$D_4$ and $C_{4v}$ NCSs exhibit, besides the global symmetries TRS and PHS, also rotation and reflection symmetries.
Two of these crystalline symmetries play an important role for the protection of zero-energy vortex-bound states. Let us
discuss these in more detail. 
We find that the $D_4$ NCS is invariant under $\pi$ rotation along the $x$ axis, which acts on the Hamiltonian~\eqref{Eq:BdG} 
as
\begin{eqnarray} \label{piRotSym}
U^{\dag}_{{\rm R}_\pi} \mathcal{H} ( {\rm R}_{\pi}  {\bf k} ) U^{\phantom{\dag}}_{{\rm R}_{\pi}} = \mathcal{H} ( {\bf k} ) ,
\end{eqnarray}
where $U_{{\rm R}_{\pi}} = \mathrm{diag} ( u^{\phantom{*}}_{{\rm R}_\pi}, u_{{\rm R}_\pi}^{\ast} )$ and
$u_{{\rm R}_\pi}$ is the spinor representation of the rotation ${\rm R}_\pi =  \mathrm{diag}  ( 1, -1, -1)$, i.e.,
$u_{{\rm R}_\pi} = \exp \left[ - i ( \pi /2) \sigma_1 \right] = - i \sigma_1 $.
The $C_{4v}$ point-group NCS, on the other hand,  satisfies the reflection symmetry 
\begin{eqnarray} \label{reflectionSymOp}
U^{\dag}_{ {\rm R}_{y}} \mathcal{H} ( {\rm R}_y {\bf k} )  U^{\phantom{\dag}}_{ {\rm R}_{y}} = \mathcal{H}( {\bf k} ),
\end{eqnarray}
with $U^{\dag}_{ {\rm R}_{y}} =  \mathrm{diag} ( u^{\phantom{*}}_{{\rm R}_y}, u_{{\rm R}_y}^{\ast} )$  and $u^{\phantom{*}}_{{\rm R}_y} = i \sigma_2 $ the spinor representation of the reflection operator 
${\rm R}_{y}=  \mathrm{diag}  ( 1, -1, 1)$.

For a pair of  vortex-antivortex lines oriented along the $z$ axis, the 
spin-singlet and spin-triplet order parameters are modified as 
\begin{align}
 \Delta_{s,t}(x,y)=\Delta_{s,t} e^{i \phi (x,y)},
 \end{align}
 where the phase angle $\phi (x,y)$
is given by $\phi(x, y)=\tan^{-1} [ 2aby/(x^2+(by)^2-a^2)]$. This describes
a vortex and antivortex line with winding number $\pm 1$ located at $(a,0)$ and $(-a,0)$, respectively.
The anisotropy of the vortex line is controlled by the parameter $b$.
In order to compute the vortex-bound states 
we set $(a,b)=(8,2)$ and
diagonalize the BdG Hamiltonian \eqref{Eq:BdG} on a 
$50\times 50\times 60$ cubic lattice with periodic boundary conditions (PBCs) in all three directions.
To simplify matters, we do not take into account the Zeeman effect. 
Most NCS topological superconductor candidate materials are extreme type-II superconductors\cite{bauerSigristBook} 
with a lower critical field $H_{c1}$ of the order of $1$~mT, corresponding to an energy scale an
order of magnitude smaller than the gap energy. For these systems, it is expected that
the Zeeman effect can be neglected for a magnetic field that is not much larger than $H_{c1}$.

The Hamiltonian in the presence of a pair of vortex-antivortex lines along the $z$ axis
breaks TRS as well as the crystal symmetries \eqref{piRotSym} and \eqref{reflectionSymOp},
but remains invariant under a combination of TRS with crystal symmetries. That is,
the $D_4$ point-group NCS with $k_z=0$ satisfies the following symmetry
\begin{subequations} \label{combSym}
\begin{equation} \label{combSymD4}
[{ \mathcal{U}_{{\rm R}_\pi}} \mathcal{U}_T ]^{-1}\widetilde{\mathcal{H}}  (x,x',y,y' ) [{ \mathcal{U}_{ {\rm R}_\pi} } \mathcal{U}_T ]  
= \widetilde{\mathcal{H}}^*(x,x',y,y' ),
\end{equation}
where
$\widetilde{\mathcal{H}} (x,x',y,y' )$ denotes the Fourier transform of $\mathcal{H} ( k_x, k_y, k_z = 0)$, 
${ \mathcal{U}_{{\rm R}_\pi}}=\delta_{x,x}\delta_{y,-y} U_{{\rm R}_\pi}$ is
 the real-space $\pi$-rotation operator,
and  $\mathcal{U}_T=\delta_{x,x}\delta_{y,y} \otimes U_T$ represents the time-reversal operator in  position space.
Similarly, for the $C_{4v}$ NCS in the presence of vortices, we find the following symmetry
\begin{equation} \label{combSymC4v}
[{\mathcal{U}_{{\rm R}_y} \mathcal{U}_T}]^{-1}\widetilde{\mathcal{H}}  (x,x',y,y', k_z) [{\mathcal{U}_{{\rm R}_y} \mathcal{U}_T}]  
= \widetilde{\mathcal{H}}^*(x,x',y,y', k_z),
\end{equation}
\end{subequations}
for all $k_z$, where $\mathcal{U}_{{\rm R}_y} = 
 \delta_{x,x}\delta_{y,-y} \otimes U_{{\rm R}_y}$ 
 denotes the reflection operator   in real space.
Here, the   matrix $\delta_{x,x}\delta_{y,-y}$ acts on the real-space basis,
while $U_{{\rm R}_y}$ acts on the Nambu basis.
We note that the  $C_2$ NCSs does not possess any symmetry of the form of Eqs.~\eqref{combSym}.
In Sec.~\ref{sec:FOUR} it is shown that symmetries~\eqref{combSym}  together with PHS lead to the protection of zero-energy vortex-bound states.

\section{Phase diagram and topological invariants}
\begin{figure}[t!]
\centering
\includegraphics[height=7 cm] {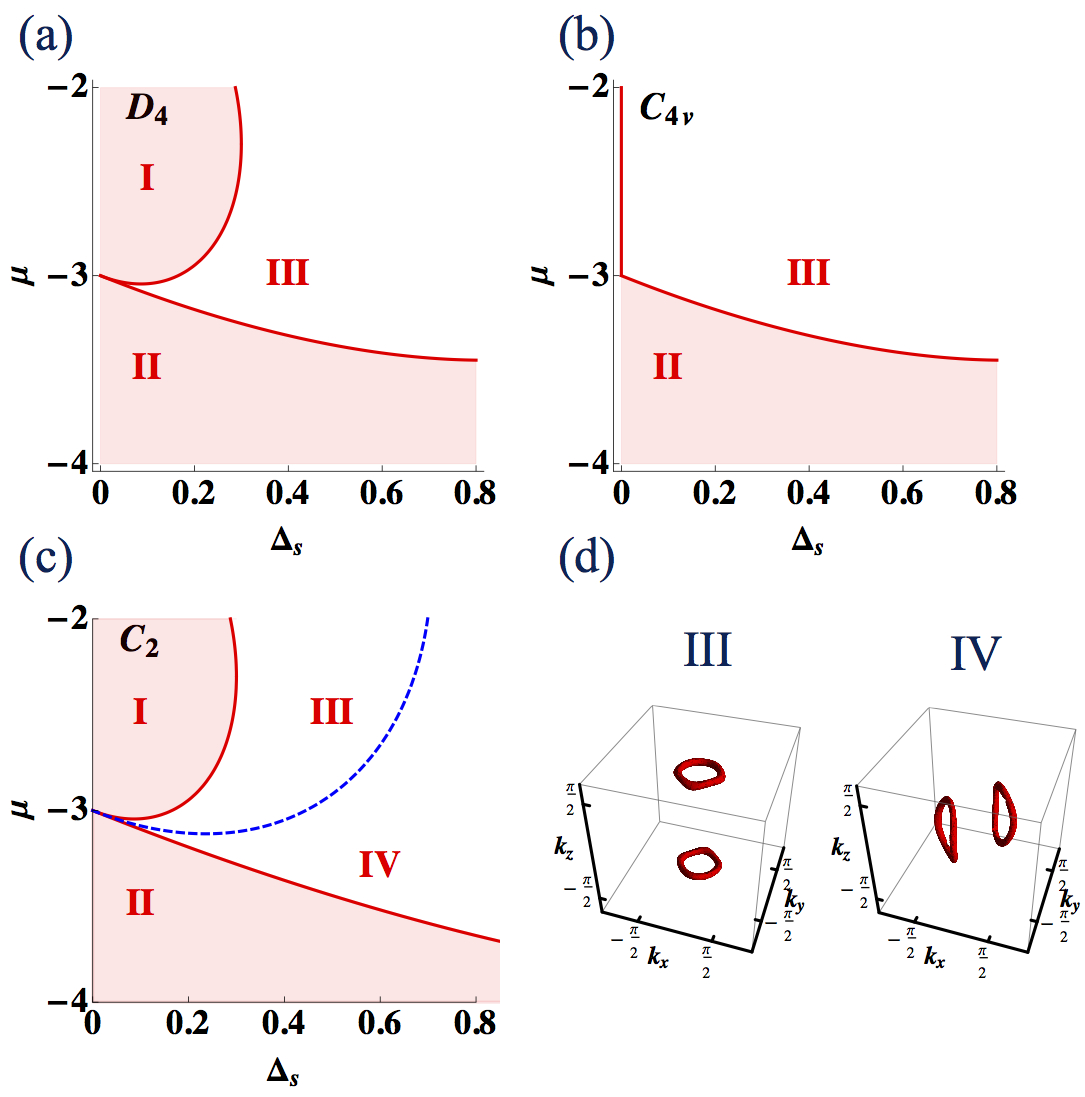}
\caption{(Color online). 
Phase diagram of Hamiltonian~\eqref{Eq:BdG} as a function of
spin-singlet pairing amplitude $\Delta_s$ and chemical potential $\mu$
for the point group (a) $D_4$ with $(a_1, a_2) = (1.0,0.3)$, (b) $C_{4v}$ with $a_1 = 1.0$, 
and (c) $C_2$ with $(a_1 = a_2 , a_3, a_4=a_5) = ( 1.0, 0.3, 0.5)$.
The fully gapped phases (red shaded regions) are characterized by the winding number $\nu_3$,
where $\nu_3=-1$ (phase I) corresponds to the topological phase, while 
$\nu_3=0$ (phase II) is  the trivial phase. The blue dashed line in panel (c) represents the phase boundary between 
the two distinct nodal
structures III and IV shown in panel (d). }
\label{fig:phase_diagram}
\end{figure}

The phase diagram of Hamiltonian~\eqref{Eq:BdG} in the absence of vortices 
is shown in Figs.~\ref{fig:phase_diagram}(a)-\ref{fig:phase_diagram}(c)
as a function of  spin-singlet pairing amplitude $\Delta_s$ and  chemical potential $\mu$.
Two fully gapped phases with trivial and nontrivial topology (phases I and II in Fig.~\ref{fig:phase_diagram}) are separated by 
a nodal superconducting phase (phases III and IV in Fig.~\ref{fig:phase_diagram}) \cite{gapclosing}.
Interestingly, for the $C_2$ point-group NCS we find that there are two distinct gapless phases
with a Lifshitz transition in between, at which the nodal rings touch each other and reconnect
in a different manner [see Figs.~\ref{fig:phase_diagram}(d) and \ref{fig: Scenario B}(a)-\ref{fig: Scenario B}(d)].
The topological properties of  the fully gapped phases I and II in Fig.~\ref{fig:phase_diagram} are characterized 
by the 3D winding number $\nu_3$,\cite{Schnyder}
which is defined as 
\begin{equation}
\nu_3=\int_{\mathrm{BZ}}\frac{d^3k}{24 \pi^2} \epsilon^{\mu\nu\rho}\mathrm{Tr}[(q^{-1} \partial_{k_\mu} q)(q^{-1} \partial_{k_\nu} q)(q^{-1} \partial_{k_\rho} q)],
\end{equation}
where $q$ is the off-diagonal block of 
the spectral projector, see Appendix~\ref{appA}.
We find that phase I is topologically nontrivial 
with $\nu_3=-1$, while
phase II is  trivial with $\nu_3=0$.
Note that this 3D winding number $\nu_3$  
is only well defined for fully gapped phases.
The topological characteristics of the nodal phases III and IV, however,
can be described by the one-dimensional winding number\cite{Schnyder,Matsuura}
\begin{equation}
\nu_1
=\frac{i}{2\pi} \oint_\mathcal{L} dk_\mu \mathrm{Tr} [q^{-1} \partial_{k_\mu} q],
\end{equation}
where $\mathcal{L}$ is a closed path that interlinks with a line node.
In both nodal phases III and IV, the winding number $\nu_1$
 evaluates to $\pm 1$ for each nodal ring.
To characterize the nodal phases it is also possible to define a one- or two-dimensional
$\mathbb{Z}_2$ topological invariant\cite{Schnyder,Matsuura} 
\begin{align} \label{Z2number}
W_{\mathcal{A}}   = 
\prod_{{\bf K}} 
\frac{\mathrm{Pf}\, \left[q^T({\bf K})\right] }
{\sqrt{ \det \left[ q ( {\bf K} ) \right] }},
\end{align}
where $\mathcal{A}$ is a time-reversal invariant line or plane embedded in the
3D Billouin zone (BZ)
 and ${\bf K}$ denotes 
the two (four) time-reversal invariant momenta on the line (plane) $\mathcal{A}$.

\section{Vortex-bound states and surface states}
\label{sec:FOUR}

In this section we discuss the surface states and vortex-bound states of nodal NCS
with the thee different point-group symmetries $D_4$, $C_{4v}$, and $C_2$.

\subsection{$D_4$ point-group NCSs}
\label{sec:FOURA}

\begin{figure}[t!]
\centering
\includegraphics[height = 5.3 cm] {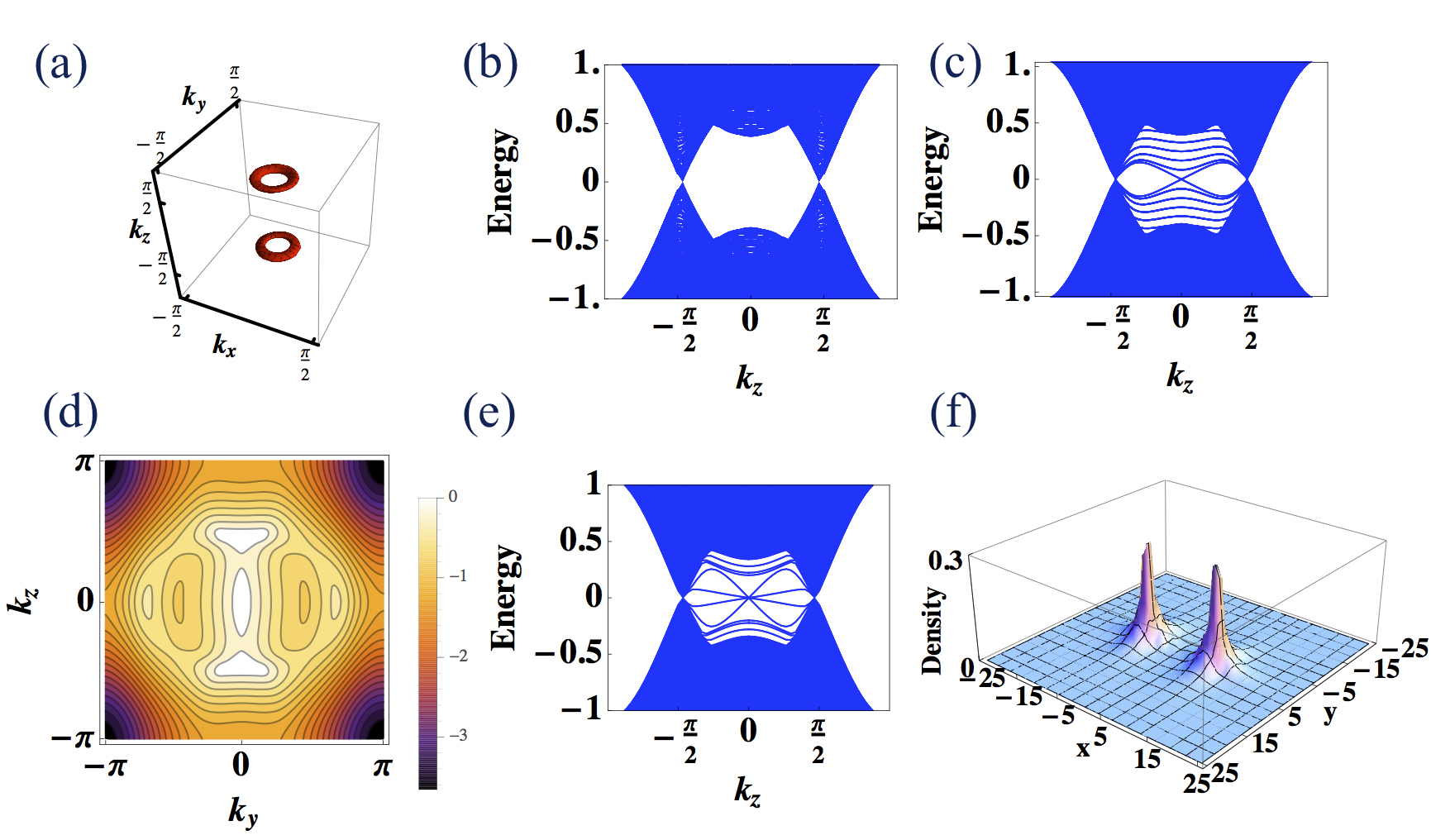}
\caption{(Color online).
Zero-energy vortex-bound states and surface states in a D$_4$ point-group NCS with $(a_1, a_2)=(1.0, 0.3)$,
 $\mu = -2.5$, and $\Delta_s = 0.5$. This parameter choice corresponds to phase III in Fig.~\ref{fig:phase_diagram}(a).
Energies are measured in the unit of hopping.
 (a)~Bulk nodal structure. 
(b) and (c): Energy spectrum in the absence of vortices as a function of $k_z$ with (b) PBCs
in all three directions and (c) OBCs along the $x$ axis but PBCs in the other two directions.
(d) Energy dispersion of the 
highest negative-energy state of the $D_4$ NCS in a (100) slab geometry. The color scale is such that
white  represents zero energy.
(e) Energy spectrum as a function of $k_z$ in the presence of a
vortex-antivortex  pair oriented along the $z$ axis. The subgap
states are localized at the vortex cores.
(f) Probability distribution of the zero-energy vortex-bound states as
a function of lattice position.}
\label{fig: Scenario AI}
\end{figure}

We start by considering a nodal NCS with $D_4$ point-group symmetry
in phase III of Fig.~\ref{fig:phase_diagram}(a).
In this region of parameter space
the bulk Bogoliubov quasiparticle spectrum
exhibits two topologically stable nodal rings, which are centered about the (001) axis [Figs.~\ref{fig: Scenario AI}(a) and~\ref{fig: Scenario AI}(b)].
The one-dimensional winding number $\nu_1$ (topological charge) of these two nodal rings is $\nu_1 = \pm 1$,
which by the bulk-boundary correspondence results in the appearance of flat-band surface states \cite{Schnyder,Schnyder1}.
In addition to the surface flat bands, nodal $D_4$ NCSs  exhibit $\mathbb{Z}_2$ Majorana surface states.
This is shown in Figs.~\ref{fig: Scenario AI}(c) and~\ref{fig: Scenario AI}(d) for the
(100) surface, where a helical Majorana cone appears at  $(k_y,k_z)=(0, 0)$ of the surface BZ.
As shown in Ref.~\onlinecite{Schnyder}, this Majorana surface state is protected by the 
one-dimensional $\mathbb{Z}_2$ topological invariant~\eqref{Z2number}
with $\mathcal{A}$ a time-reversal invariant line.
Choosing $\mathcal{A}$ to be oriented along the $k_x$ axis with $(k_y, k_z)$ held fixed,
we find that 
$W_{\mathcal{A}} = -1$ at $(k_y, k_z)=(0, 0)$, which signals the appearance of a zero-energy helical Majorana state.
At the other three time-reversal invariant momenta of the surface BZ there are no surface states, in agreement with $W_{\mathcal{A}} = +1$ 
for theses surface momenta.

 $D_4$ point-group NCSs support zero-energy
helical Majorana states not only on the surface but also inside vortex cores.
This is illustrated in Fig.~\ref{fig: Scenario AI}(e), which shows
the energy spectrum in the presence of a pair of vortex and antivortex lines oriented along
the $z$ axis. At energies smaller than the bulk gap there appear
vortex-bound states which disperse linearly along the vortex lines. 
These vortex-bound states are similar to the one-dimensional 
helical Majorana modes that exist
at the edge of a fully gapped topological NCS in two dimensions. 
The numerical simulations of Fig.~\ref{fig: Scenario AI} are in excellent
agreement with an analytical derivation of the vortex-bound states,\cite{continuum}
cf.\ Appendix~\ref{App: Continuous model} and Ref.~\onlinecite{Lu}. 

The zero-energy vortex-bound states at $k_z=0$ are protected by a combination of
$\pi$ rotation, time-reversal, and particle-hole symmetry; see Eq.~\eqref{combSymD4}.\cite{FangPRL14,MizushimaPRL12,KenArXiv} 
Namely, we find that these zero energy modes
 are eigenstates of the chiral operator $\mathcal{S} =\mathcal{U}_{{\rm R}_\pi}  \mathcal{U}_T \mathcal{U}_P $
 and their stability is guaranteed by the conservation of chiral symmetry. That is,
 the doubly degenerate zero energy states at the vortex core are eigenstates of $\mathcal{S}$ with eigenvalue $+1$, whereas
the two zero-energy modes at the anti-vortex have eigenvalues $-1$.
Without breaking chiral symmetry, a zero-energy state in the vortex core with chirality eigenvalue $+1$ can only be removed together with 
a zero mode at the anti-vortex with chirality $-1$. Hence, in the limit where vortex and anti-vortex cores are separated by a large distance,
the zero-energy vortex-bound states are robust against any local perturbation that does not break chiral symmetry $\mathcal{S}$. In particular, 
the  zero-energy vortex-bound states remain unperturbed by
the chiral symmetric Zeeman field $h_z \sigma_z \otimes \sigma_z$.

By employing  continuous deformations of the quasiparticle spectrum of Hamiltonian~\eqref{Eq:BdG},
one can show that the $\mathbb{Z}_2$ surface states and the helical vortex-bound states of the nodal NCS
with $D_4$ point-group symmetry [phase III in Fig.~\ref{fig:phase_diagram}(a)] originate from the nontrivial properties of
the adjacent fully gapped phase of the $D_4$ NCS [phase I in Fig.~\ref{fig:phase_diagram}(a)].
To be more specific, let us fix the momentum along the vortex line (e.g., to $k_z =0$) 
and consider adiabatic deformations connecting phase III to phase I that do not close 
the bulk gap at this particular momentum. During this  deformation process, the two nodal
 rings shrink to nodal points at the north and south poles of the Fermi sphere and vanish,
while the zero-energy vortex and surface states at $k_z=0$ remain unaffected. 
Moreover, the $\mathbb{Z}_2$ invariant $W_{\mathcal{A}}$ of the nodal phase III can be shown to be directly
related to the 3D winding number $\nu_3$ of the fully gapped phase I (cf.~Ref.\ \onlinecite{Sato2}).
Hence,  the zero-energy vortex and $\mathbb{Z}_2$ surface states of a nodal  $D_4$  NCS
are adiabatically connected to the vortex and surface states of a fully gapped topological NCS with 
 $D_4$  point-group symmetry.
A similar deformation process connecting phase~III to phase~II, on the other hand,  
does not exist, since upon crossing the transition line between phase~III and phase~I, the
 nodal rings approach each other and pair-annihilate. As a result, the zero-energy
$\mathbb{Z}_2$ surface states and vortex-bound states disappear
as one traverses the transition line.

\subsection{$C_{4v}$ point-group NCSs}

\begin{figure}[t!]
\centering
\includegraphics[height = 5.3 cm] {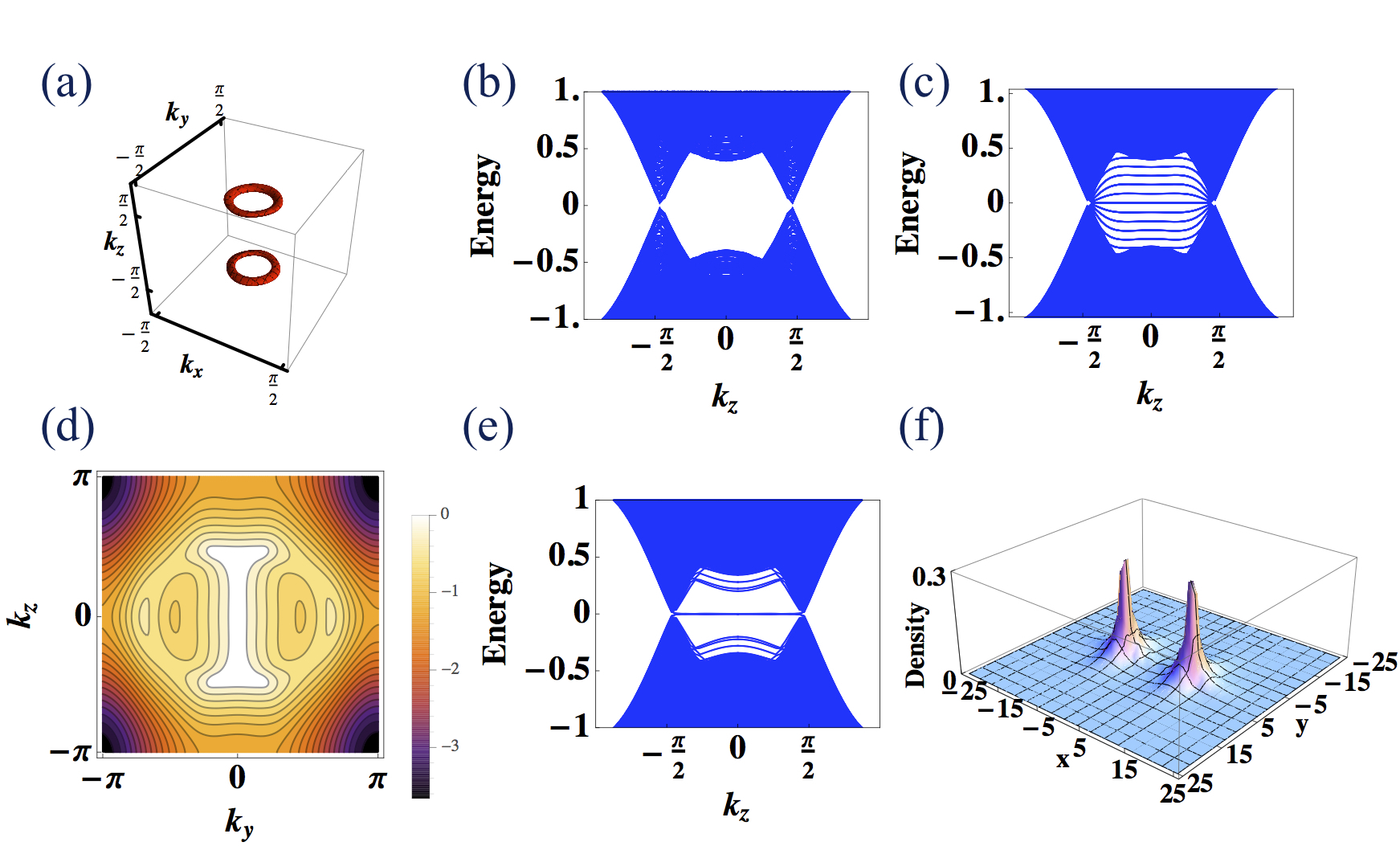}
\caption{(Color online).
Same as Fig.~\ref{fig: Scenario AI} but for a $C_{4v}$ point-group NCS with $a_1 = 1.0$,
$\mu=-2.5$, and $\Delta_s =0.5$, corresponding to phase~III in Fig.~\ref{fig:phase_diagram}(b). }
\label{fig: Scenario AII}
\end{figure}

Next we study surface and vortex-bound states of a nodal $C_{4v}$ point-group NCS
in phase III of Fig.~\ref{fig:phase_diagram}(b).
The bulk quasiparticle spectrum in this nodal phase [Figs.~\ref{fig: Scenario AII}(a) and~\ref{fig: Scenario AII}(b)] resembles
the one of the $D_4$ NCS [Figs.~\ref{fig: Scenario AI}(a) and~\ref{fig: Scenario AI}(b)] and shows two nodal rings around the poles of the 
Fermi sphere. 
These line nodes have a nontrivial topological charge, which, as a consequence of the bulk-boundary correspondence,
lead to the appearance of flat-band surface states. In addition,  $C_{4v}$ NCSs  support 
helical arc surface states, that connect  the projected nodal rings in the surface BZ [see Figs.~\ref{fig: Scenario AII}(c) and \ref{fig: Scenario AII}(d)].
These helical arc surface states are protected by a two-dimensional $\mathbb{Z}_2$ number,
which is defined for each plane perpendicular to the (001) direction, (i.e., for planes with fixed~$k_z$) \cite{Schnyder1}, see Appendix~\ref{App: Higher order}.
The arc surface states of  $C_{4v}$ NCSs can be viewed as superconducting analogues of the Fermi arcs in time-revesal symmetric Weyl semimetals~\cite{Morimoto, Ojanen, footnote_Weyl},
or alternatively as time-reversal invariant versions of the arc states in the A phase of superfluid $^3$He \cite{volovik11,volovikBOOKS}.

Due to the bulk-vortex correspondence \cite{volovik11}, vortex lines in $C_{4v}$ NCSs support zero-energy bound states 
which are dispersionless along the vortex line.
This is illustrated in Figs.~\ref{fig: Scenario AII}(e) and ~\ref{fig: Scenario AII}(f)
for a pair of vortex and antivortex lines that are oriented along the $z$ axis. Just as the arc surface states, these flat-band
vortex-bound states connect the projected bulk nodes in $k_z$ momentum space.
Following similar arguments as in Sec.~\ref{sec:FOURA}, it can be shown that the zero-energy
vortex-bound states of the $C_{4v}$ NCSs for any fixed $k_z$
are protected by the chiral symmetry $\mathcal{S} =\mathcal{U}_{{\rm R}_y}  \mathcal{U}_T \mathcal{U}_P$; see Eq.~\eqref{combSymC4v}.
Using an adiabatic deformation of the Bogoliubov quasiparticle spectrum that does not close the bulk energy gap
at the momenta $k_z$ in between the two projected nodal rings, we find that the flat-band vortex states
and the arc surface states of the $C_{4v}$ NCSs can be related to the vortex states and surface states of
a time-reversal symmetric Weyl superconductor. That is, upon approaching the boundary
of phase III in Fig.~\ref{fig:phase_diagram}(b) where $\Delta_s=0$ and $\mu > -3$,  the nodal rings shrink to points at 
the north and south poles of the Fermi sphere and the $C_{4v}$ NCSs turns into a time-reversal invariant Weyl superconductor, 
i.e., a time-reversal symmetric analog of the A phase of ${}^3$He.

\begin{figure}[t!]
\centering
\includegraphics[height = 5.5 cm] {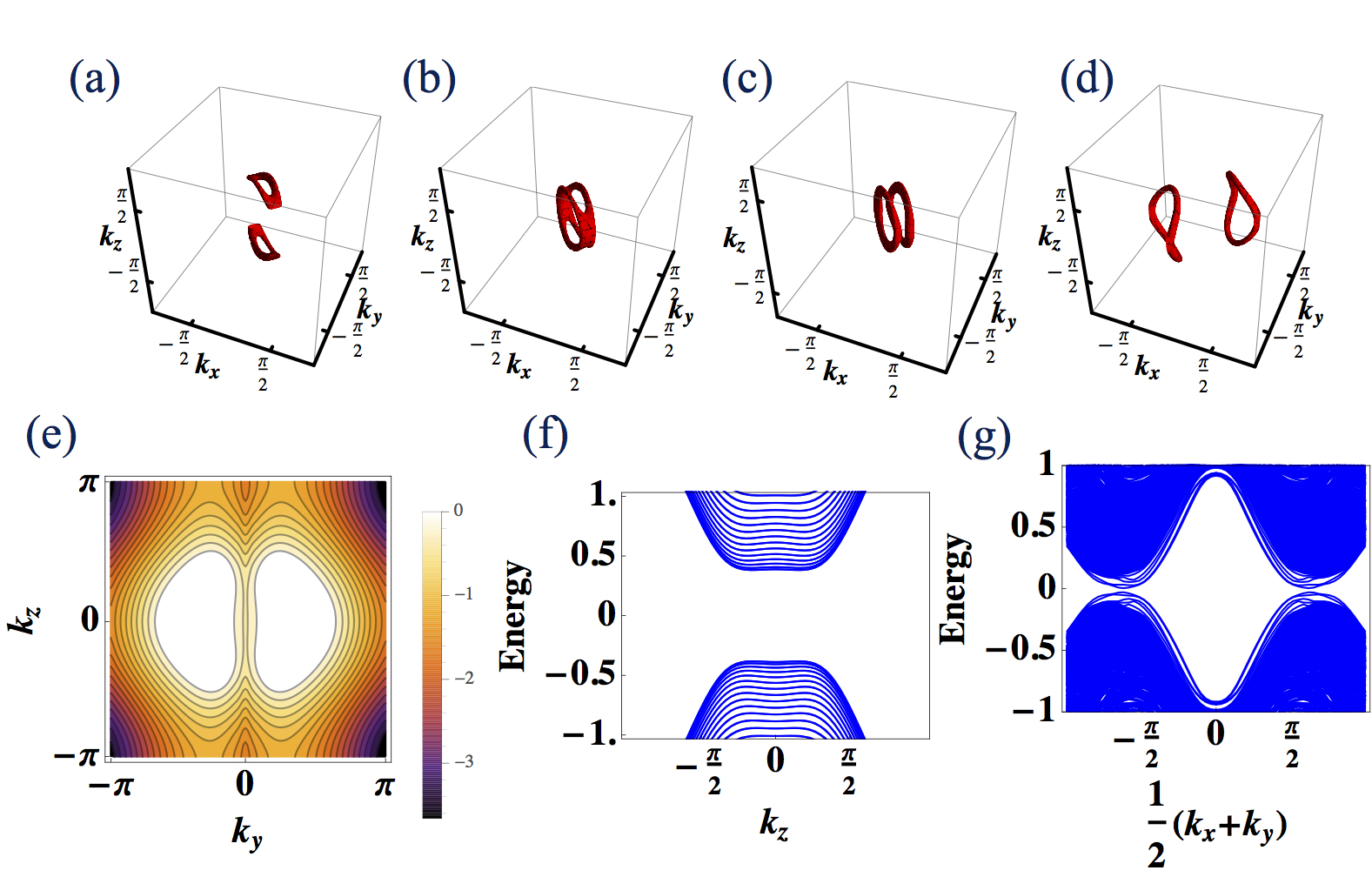}
\caption{(Color online).
Bulk nodal structure, surface states, and finite-energy vortex-bound states
for a $C_2$ point-group NCS with $a_1=a_2=1.0$, $a_3=0.3$, and $a_4=a_5=0.5$.
Energies are measured in the unit of hopping.
(a)-(d): Evolution of the
 bulk nodal structure as one moves along a path in the $(\mu , \Delta_s)$ phase 
diagram of Fig.~\ref{fig:phase_diagram}
from phase~III [panels (a) and (b) with $(\mu,\Delta_s)=(-2.5,0.5)$ and $(-2.9,0.5)$, respectively] to phase~IV 
[panels (c) and (d) with $(\mu,\Delta_s)=(-3.1,0.5)$ and $(-3.1,1.5)$, respectively].
(e) and (f): Energy spectrum in the absence of vortices for (e) the highest negative-energy state 
and (f) all the states with $k_y=0$
of a $C_2$ NCS in a (100) slab geometry with the same parameters as in panel (d).
(g) Energy spectrum in the presence of a vortex-antivortex pair oriented along the 
(110) direction as a function of momentum parallel to the vortex lines, $k_{\parallel} = \frac{1}{2}(k_x+k_y)$,
with the same parameters as in panel (d). The subgap states are localized at the vortex cores. }
\label{fig: Scenario B}
\end{figure}

\subsection{$C_{2}$ point-group NCSs}

Lastly, we examine the surface and vortex-bound states of NCSs with  $C_{2}$ point-group symmetry.
The phase diagram of  $C_2$ NCSs as a function of spin-singlet pairing amplitude $\Delta_s$ and chemical
potential $\mu$ displays two distinct nodal phases, which differ in the orientation of the nodal rings 
[Figs.~\ref{fig:phase_diagram}(c), \ref{fig:phase_diagram}(d), and \ref{fig: Scenario B}(a)-\ref{fig: Scenario B}(d)].
In phase III the nodal rings are oriented along the (001) axis, while in phase IV they are
centered about the (110) direction.
As in the previous two cases, the topological characteristics of these nodal rings, which is described
by the one-dimension winding number $\nu_1$, leads to the appearance of flat-band surface states.
In addition, phase III supports $\mathbb{Z}_2$ Majorana surface states, whereas phase IV does not
exhibit any additional surface states. 
This is exemplified in Figs.~\ref{fig: Scenario B}(e) and~\ref{fig: Scenario B}(f), which show
the energy spectrum at the (100) surface of a $C_2$ NCS in phase IV. 
Flat-band surface states appear within regions of the surface BZ that are bounded by
the projected bulk nodal rings. But otherwise there exist no additional surface states in phase IV.
Indeed, the energy spectrum along the $k_z=0$ line is fully gapped [Fig.~\ref{fig: Scenario B}(f)].
Using the same adiabatic deformations as before, we find that phase III can be connected
to phase I, showing that the $\mathbb{Z}_2$ surface states of phase III originate from
the topological properties of the fully gapped phase I.
Phase IV, on the other hand, can be deformed into phase II
by shrinking the nodal rings into points at opposite sides of the Fermi surface until 
they vanish, which 
corroborates our finding that there are no additional surface states in phase IV.

In contrast to NCSs with $D_4$ or $C_{4v}$ point-group symmetry, 
NCSs with monoclinic point-group  $C_{2}$ do not support any zero-energy vortex-bound states,
neither in the fully gapped phases~I and II nor in the nodal phases III and IV.
This is in line with our finding that
the  chiral symmetry $\mathcal{S}$ (i.e., the combination of 
reflection ($\pi$ rotation), 
particle-hole, and time-reversal symmetry),
which is present for  $D_4$ and $C_{4v}$ NCSs but absent for  $C_2$ NCSs,
guarantees the
the stability of the zero-energy vortex-bound states.
The absence of zero-energy vortex states in $C_{2}$ point-group NCSs is demonstrated in Fig.~\ref{fig: Scenario B}(g) for phase IV, which
shows the energy spectrum for a vortex-antivortex pair oriented along the (110) axis, 
and also follows from an analytical argument,\cite{footnote_bdg_cont2} see Appendix~\ref{App: Continuous model}.

\section{Summary and discussion}
In summary, we have studied  zero-energy vortex-bound states in 3D nodal and fully gapped NCSs. 
While vortex lines in NCSs with tetragonal point-group $D_4$ and $C_{4v}$  support zero-energy vortex-bound states,
$C_2$ point-group NCSs do not exhibit any Majorana vortex-bound states. We have found that
the existence of Majorana vortex-bound states in nodal NCSs correlates 
with the appearance of Majorana cone and arc surface states.
The zero-energy vortex states in $C_{4v}$ ($D_4$) NCSs are protected 
by a combination of reflection ($\pi$ rotation), time-reversal, and particle-hole symmetry, see Eq.~\eqref{combSym}.
This is reminiscent of
the zero modes at dislocation lines of band-topological insulators which are
stabilized by certain space group symmetries~\cite{Slager1}.

Our findings have implications for experiments on 3D NCSs and on heterostructures,
in which topological superconductivity is induced via the proximity effect of a conventional $s$-wave superconductor \cite{Sasaki,zhangeKaneMele}.  
 Vortex-bound states can be directly observed
in ordinary and spin-resolved scanning tunneling microscopy \cite{SunArXiv14}. The
 helical Majorana vortex states of $D_4$ point-group NCSs  can carry currents along the vortex lines,
which could  in principle  be detected using thermal transport measurements~\cite{Kashyap}. 
Moreover, the vortex-bound states are expected to be observable in terms of the 
cross-correlated responses
between the orbital angular momentum ${\bf L}$ and the thermal polarization ${\bf P}_E$ of a 3D topological SC,
which were recently discussed in Ref.~\onlinecite{Nomura}.
The so-called gravitomagnetoelectric polarizability of a 3D topological SC  
(i.e., the   analog of the magnetoelectric
polarizability  of a 3D topological insulator)
is given by
\begin{align} \label{GMEE}
\chi^{ab}_g
=
\frac{\partial L^a}{\partial E^b_{g}}
=
\frac{\partial P^a_E}{\partial \Omega^b},
\quad
a,b=x,y,z, 
\end{align}
where 
${\bf \Omega}$ is the (external) angular velocity of the SC and
${\bf E}_g = -T^{-1} \pmb{ \nabla} T$ represents the temperature gradient.
Note that the thermal polarization ${\bf P}_E$ is related to the distribution of the induced heat $Q$ via  
$\Delta Q= - \pmb{\nabla}\cdot {\bf P}_E$.
According to Eq.~\eqref{GMEE}  a thermal polarization (entropy polarization) $P_E^{a}$ can be generated 
by rotating the system with angular velocity $\Omega^{b}$.
The presence of vortex lines leads to an additional contribution to the angular momentum
and hence to an additional accumulation of entropy (heat) at the top and bottom surfaces of the 3D SC.
Vortex-bound states, on the other hand, can carry a thermal current connecting top and bottom surfaces.

\section{Acknowledgments} We thank 
J. C. Y. Teo, P. M. R. Brydon, and C. Timm for useful discussions.
SR is supported by Alfred P. Sloan Research Fellowship (FG-BR2014-029).


\appendix

\section {$q$ matrix}
\label{App:$q$ matrix}
\label{appA}

A Hamiltonian $H =  \sum_{ {\bf k} } \Psi^{\dag}_{\bf k}   \mathcal{H}({\bf k } ) \Psi_{\bf k} $
which preserves chiral symmetry $S$ can be brought into
block off-diagonal form, 
\begin{eqnarray}
 \label{eqChrialBasis}
\tilde{\mathcal{H}} ( {\bf k} )=V\mathcal{H} ({\bf k}) V^\dagger
=
\left(\begin{array}{cc}
0 & D ( {\bf k} ) \\
D^{\dag} ( {\bf k} ) & 0 \\
\end{array}\right),
\end{eqnarray} 
where $V$ is a unitary transformation that diagonalizes the chiral symmetry operator $S$. 
In general, one can always deform a Hamiltonian into a flat-band Hamiltonian without altering its topological
features. The flat-band Hamiltonian $Q({\bf k})$ of Hamiltonian $\tilde{\mathcal{H}} ( {\bf k} )$ can be defined in terms of the spectral projector $P({\bf k})$ 
\begin{eqnarray}
\label{eq:projector}
Q({\bf k})
&=&
\mathbb{I}_{4N}
-
2 P ({\bf k})  \notag\\
&=&
\mathbb{I}_{4N} 
- 2 \sum_{ a =1 }^{2 N}  
\left( \begin{array}{c}
\lambda^-_a ({\bf k}) \\
\mu^-_a ({\bf k}) \\
\end{array} \right)
\left( \begin{array}{cc}
\left[ \lambda^-_a ({\bf k}) \right]^{\dag} &
\left[ \mu^-_a ({\bf k}) \right]^{\dag} 
\end{array} \right) ,  \nonumber\\
\end{eqnarray}
where $\big( \begin{array}{c c} \lambda^-_a ({\bf k}) & \mu^-_a ({\bf k}) \end{array} \big)^\mathrm{T}$ are the negative-energy eigenfunctions of $\tilde{\mathcal{H}} ( {\bf k}) $,
which are obtained from the eigenequation
\begin{equation} \label{eigenEQ}
\left(
\begin{array}{cc}
0 & D ( {\bf k} ) \\
D^{\dag} ( {\bf k} ) & 0 
\end{array}
\right)
\left(
  \begin{array}{cc}
\lambda^{\pm}_a ( {\bf k} ) \cr
\mu^{\pm}_a ( {\bf k} ) \cr
\end{array}
\right)
=
\pm E_a ( {\bf k} ) 
\left(
  \begin{array}{cc}
\lambda^{\pm}_a ( {\bf k} ) \cr
\mu^{\pm}_a ( {\bf k} ) \cr
\end{array}
\right).
\end{equation}
Here, 
$a=1,\ldots,2N$
denotes the combined band and spin index (we consider $N$ bands and two spin degrees of freedom).
The eigenstates $(\lambda^{\pm}_a ( {\bf k} ),\mu^{\pm}_a ( {\bf k} ) )^{\mathrm{T}}$
can be obtained from the eigenstates of $D( {\bf k} )D^{\dagger}( {\bf k} )$ or $D^{\dagger}( {\bf k} )D( {\bf k} )$,
\begin{align}
&D( {\bf k} )D^{\dagger}( {\bf k} ) \lambda_a( {\bf k} ) = E_a( {\bf k} )^2 \lambda_a( {\bf k} ), \\
&D^{\dagger}( {\bf k} ) D( {\bf k} ) \mu_a ( {\bf k} )= E_a( {\bf k} )^2 \mu_a ( {\bf k} ),
\end{align}
where the eigenstates $\lambda_a( {\bf k} )$ and $\mu_a( {\bf k} )$ are normalized to be $1$, i.e., 
$\lambda_a^\dagger( {\bf k} ) \lambda_a( {\bf k} ) = \mu^\dagger_a ( {\bf k} )\mu_a( {\bf k} )=1$.
With this, we find that the eigenstates of  $\tilde{\mathcal{H}}(\mathbf{k})$ are
$(\lambda^{\pm}_a ( {\bf k} ),\mu^{\pm}_a ( {\bf k} ) )^{\mathrm{T}} = \frac{1}{\sqrt{2}}(\lambda_a ( {\bf k} ),\pm \mu_a ( {\bf k} ) )^{\mathrm{T}} $.
Hence the flat-band Hamiltonian $Q(\mathbf{k})$ defined from the spectral project $P(\mathbf{k})$
holds this off-diagonal form, 
\begin{eqnarray}
\label{eqQtmp}
Q({\bf k}) &= \mathbb{I}_{4N} - 2 P( {\bf k} ) =
\left(
\begin{array}{cc}
0 & q({\bf k})\\
q^{\dag}({\bf k})& 0
\end{array}
\right),
\end{eqnarray}
where $q(\mathbf{k})=\sum_a \lambda_a(\mathbf{k}) \mu^\dagger_a (\mathbf{k})$, for more details; see Refs.~\onlinecite{Schnyder,Matsuura}.

\section{Continuum BdG equations}
\label{App: Continuous model}

The continuum BdG equation can be expressed as 
\begin{align}
\mathcal{H} \psi=& \left(
\begin{array}{cc}
h & \Delta\\
\Delta^{\dagger}& -h^*
\end{array}
\right) \psi = E \psi,
\end{align}
where $h= (-\frac{\nabla^2}{2m}-\mu) \mathbb{I}_{2 \times 2 }+\alpha {\bf l}( {\bf k}) \cdot \pmb{\sigma}$, with $m$   the effective mass, $\mu$   the chemical potential, 
and $\alpha {\bf l}( {\bf k}) \cdot \pmb{\sigma}$ the Rashba-type SOC with strength $\alpha$.
The pairing term has the form $\Delta=(\Delta_s+\frac{1}{2}\pmb{\nabla}\cdot {\bf D} + {\bf D} \cdot \pmb{\nabla}) (i \sigma_2)$,
where $\Delta_s$ is the singlet pairing amplitude and ${\bf D}= - i \pmb{\nabla}_{\bf k} (\Delta_p {\bf l}({\bf k}) \cdot \pmb{\sigma})$ presents the triplet pairing. Here, $\pmb{\sigma}=(\sigma_1, \sigma_2, \sigma_3)$
is the vector of Pauli matrices.
Without loss of generality,
we consider the spin-orbit coupling vector ${\bf l}({\bf k})=(a_1 k_x +a_4 k_y, a_1 k_y +a_4 k_x, a_3 k_z)$.
A vortex line along the $z$ direction can be introduced by adding a phase on both singlet and triplet pairing amplitudes,
$\Delta_s \to e^{i \theta}\Delta_s$  and $\Delta_t \to e^{i \theta}\Delta_t$. 
In the cylindrical coordinate, the normal-state Hamiltonian and the pairing term are

\begin{widetext}
\begin{align}
h= &(-\frac{1}{2m} (\partial_r^2 +\frac{1}{r}\partial_r +\frac{1}{r^2}\partial_\theta^2 + \partial_z^2) - \mu ) \mathbb{I}_{2 \times 2} \notag\\
&+\alpha  
\left(
\begin{array}{cc}
a_3(-i \partial_z ) & -ia_1 e^{-i\theta}(\partial_r - \frac{i}{r}\partial_\theta)-i a_4 e^{i \theta} (- i \partial_r +\frac{1}{r}\partial_\theta)      \\
-ia_1 e^{i\theta}(\partial_r + \frac{i}{r}\partial_\theta)-i a_4 e^{-i \theta} ( i \partial_r +\frac{1}{r}\partial_\theta) & a_3(i \partial_z )
\end{array}
\right),    \notag\\
\Delta= &\Delta_s e^{i \theta} (i \sigma_2) \notag\\
&-i \Delta_t
\left(
\begin{array}{cc}
a_1(-\partial_r + i \frac{1}{r} \partial_\theta -\frac{1}{2r}) + i e^{2i \theta} a_4 (\partial_r + i \frac{1}{r}\partial_{\theta}-\frac{1}{2r}) & a_3 \partial_z    \\
a_3 \partial_z & e^{2i \theta}a_1(\partial_r + i \frac{1}{r} \partial_\theta -\frac{1}{2r}) + i a_4 (\partial_r - i \frac{1}{r}\partial_{\theta}+\frac{1}{2r})
\end{array}
\right).   
\end{align}
A general solution of this continuum BdG equation is 
$\psi(r, \theta, z)=e^{i k_z z} [\tilde{u}_\uparrow(r)e^{i m_1 \theta}, \tilde{u}_\downarrow(r)e^{i m_2 \theta}, \tilde{v}_\uparrow(r)e^{i n_1 \theta}, \tilde{v}_\downarrow(r)e^{i n_2 \theta}]^{\mathrm{T}}$.
The radial part of a localized solution must be of the form $f(r) \sim e^{-\kappa r}$ with $\mathrm{Re}[\kappa] > 0$.
In the asymptotic limit ($1/ r \to 0$),
we can neglect all $\frac{1}{r}$ and $\frac{1}{r^2}$ terms. We find for $(m_1,m_2, n_1, n_2)=(0,1,0,-1)$ and $a_4=0$, that the continuum
BdG equation has a localized zero-energy solution for (i) $k_z=0$ or (ii) $a_3=0$, that satisfies 

\begin{align}
\left(
\begin{array}{cccc}
-\frac{1}{2m} \kappa^2-\mu
& i \alpha a_1 \kappa e^{- i \theta} 
& -i \Delta_t a_1 \kappa
& \Delta_s e^{i \theta} \\ 
i \alpha a_1 \kappa e^{i \theta} 
& -\frac{1}{2m}\kappa^2-\mu
& -\Delta_s e^{i \theta} 
& i \Delta_t a_1 \kappa e^{2 i \theta} \\ 
i \Delta_t a_1 \kappa 
& - \Delta_s e^{- i \theta} 
& \frac{1}{2m} \kappa^2 + \mu  
& -i \alpha a_1 \kappa e^{i \theta} \\
\Delta_s e^{-i \theta} 
& - i \Delta_t a_1 \kappa e^{-2 i \theta} 
&- i \alpha a_1 \kappa e^{-i \theta} 
& \frac{1}{2m} \kappa^2 + \mu 
\end{array}
\right) \left(\begin{array}{c}\tilde{u}_{\uparrow}(r) \\\tilde{u}_{\downarrow}(r) e^{i \theta} \\\tilde{v}_{\uparrow}(r) \\\tilde{v}_{\downarrow}(r) e^{-i\theta}\end{array}\right)
=0.
\label{Eq.cBdG}
\end{align}
Note that this situation corresponds to a zero-energy bound-state solution at $k_z=0$ for the $D_4$ point-group NCS.
In addition, the decay length $\kappa$ is determined by solving the roots of 
the following determinant
\begin{align}
\mathrm{Det}\left(
\begin{array}{cccc}
-\frac{1}{2m} \kappa^2-\mu
& i \alpha a_1 \kappa 
& -i \Delta_t a_1 \kappa
& \Delta_s  \\ 
i \alpha a_1 \kappa 
& -\frac{1}{2m}\kappa^2-\mu
& -\Delta_s  
& i \Delta_t a_1 \kappa \\ 
i \Delta_t a_1 \kappa 
& - \Delta_s  
& \frac{1}{2m} \kappa^2 + \mu  
& -i \alpha a_1 \kappa \\
\Delta_s 
& - i \Delta_t a_1 \kappa 
& -i \alpha a_1 \kappa 
& \frac{1}{2m} \kappa^2 + \mu 
\end{array}
\right)=0.
\end{align}
There are only two decaying solutions ($\mathrm{Re}[\kappa]>0$)
\begin{align}
\kappa_{\pm}=\sqrt{-2 a_1^2 m^2(\alpha^2 +\Delta_t^2)-2m \mu \pm 2 \sqrt{[a_1^2 m^2 (\alpha^2 + \Delta_t^2)+ m \mu]^2-m^2(\Delta_s^2 +\mu^2)} },
\end{align}
with the condition 
\begin{align}
-a_1^2 m^2(\alpha^2 + \Delta_t^2) - m \mu - \mathrm{Re} \left[\sqrt{[a_1^2 m^2 (\alpha^2 + \Delta_t^2)+ m \mu]^2-m^2(\Delta_s^2 +\mu^2)} \right] >0.
\end{align}
\end{widetext}

\section{Higher-order SOC in NCSs with $C_{4v}$ point-group symmetry}
\label{App: Higher order}

For the $C_{4v}$ NCS with ${\bf l}( {\bf k} )$ given by Eq.~\eqref{SOCforC4v} (i.e., only the lowest order SOC term)
one finds that the BdG Hamiltonian $\mathcal{H} ( {\bf k}Ê)$ is an even function of $k_z$. 
Hence, an effective two-dimensional layer with fixed $k_z$
 satisfies time-reversal and particle-hole symmetries, $U_T^{-1}\mathcal{H}(k_x, k_y)U_T =\mathcal{H}^*(-k_x, -k_y)$
and $U_P^{-1}\mathcal{H}(k_x, k_y)U_P =-\mathcal{H}^*(-k_x, -k_y)$.
It turns out that each layer with fixed $k_z$  belongs to class DIII and the two-dimensional $\mathbb{Z}_2$ topological invariant~\eqref{Z2number} can be computed.
For $k_z$ within the two nodal rings, we find that the $\mathbb{Z}_2$ number takes on a nontrivial value, which leads to the appearance 
of   helical arc surface states. 

\begin{figure}[t!]
\centering
\includegraphics[height = 3.5 cm] {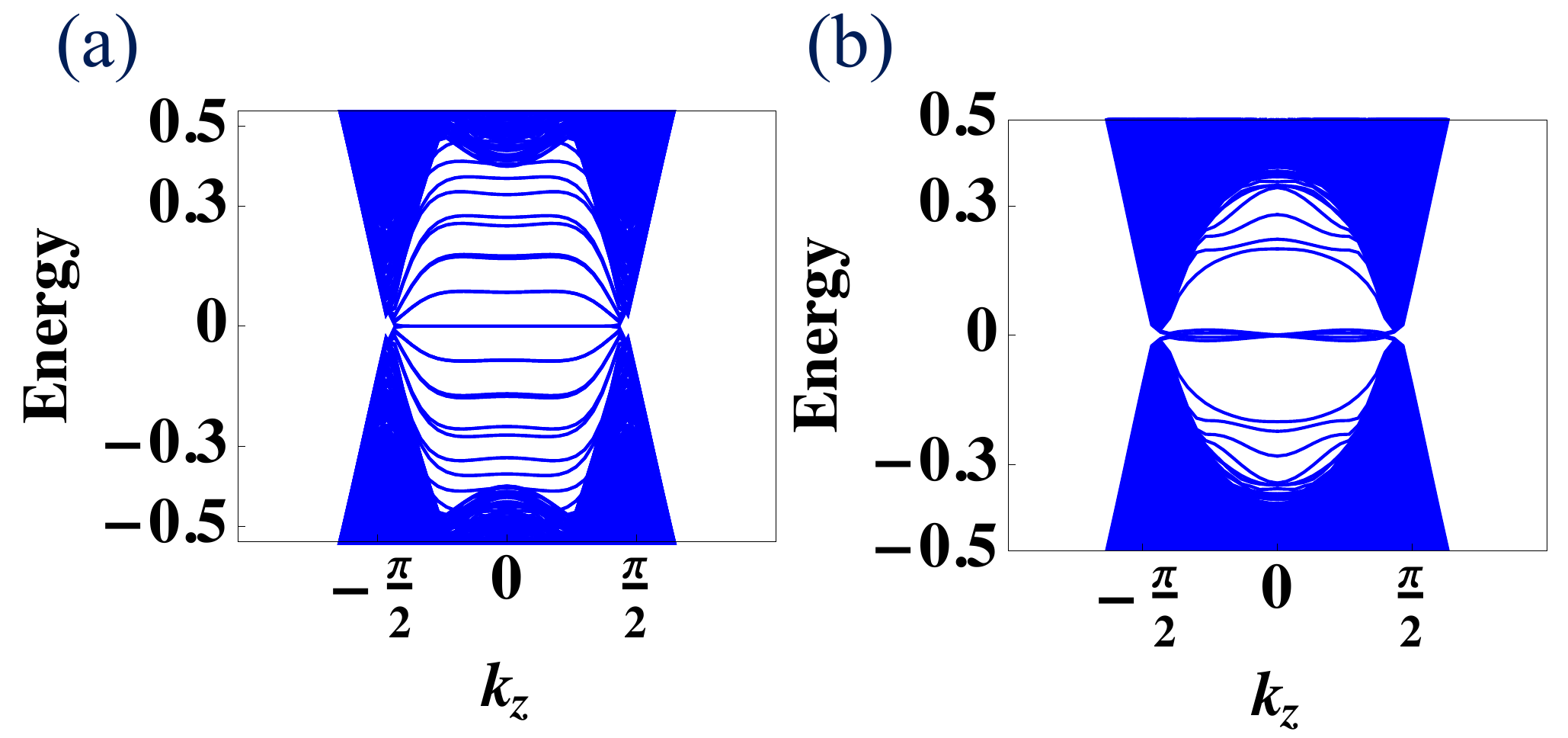}
\caption{(Color online).
Energy spectrum as a function of $k_z$ of a $C_{4v}$ NCS with $(a_1, a_2)=(1.0, 2.0 )$, $\mu=-2.5$, and $\Delta_s = 0.5$.
(a)~Without vortices and OBC along the $x$ axis but PBCs in the other two directions.
(b)~With a pair of vortex-antivortex lines
oriented along the $z$ axis and PBCs in all three directions.} 
\label{fig: D4_2nd}
\end{figure}

It is interesting to ask whether this reasoning remains valid upon inclusion of higher-order terms in the spin-orbit interaction. 
Up to second order the SOC vector $\mathbf{l}(\mathbf{k}) $ for the $C_{4v}$ point-group is given by
\begin{align}
\mathbf{l}(\mathbf{k}) =& a_1 \left[ \sin k_y  \hat{\bf e}_1  -\sin k_x \hat{\bf e}_2 \right] 
 \notag \\
&+
 a_2 \sin k_x \sin k_y \sin k_z (\cos k_x - \cos k_y) \hat{\bf e}_3 .
\nonumber
\end{align}
We observe that the second-order term is an odd function of $k_z$, and hence two-dimensional layers with fixed $k_z$ are no longer symmetric under TRS and PHS.
Our numerics shows that upon inclusion of the second-order term with $a_2 \ne 0$, the flat-band vortex-bound states become dispersive [Fig.~\ref{fig: D4_2nd}(b)].
However, the arc surface states remain unaffected [Fig.~\ref{fig: D4_2nd}(a)]. 
It turns out that the arc surface states on the (010) and (100) faces are protected by a reflection symmetry
which leaves the surface plane invariant.
The Hamiltonian in a (010) slab geometry is invariant under the following reflection symmetry:
\begin{eqnarray}
\mathcal{U}_{{\rm R}_x}^{\dag} \tilde{\mathcal {H}}(y, y'; k_x, k_z) \mathcal{U}_{{\rm R}_x}^{\phantom{\dag}}   = \tilde{\mathcal {H}}(y, y'; -k_x, k_z) ,
\end{eqnarray}
with $\mathcal{U}_{{\rm R}_x} = \delta_{y,y} \otimes [\sigma_3 \otimes \sigma_1]$, where $ \delta_{y,y}$ acts
on the real-space basis and $ \sigma_3 \otimes \sigma_1$ acts on the momentum-space Nambu basis.
The Hamiltonian in a (100) geometry, on the other hand, is invariant under the following mirror symmetry:
\begin{align}
\mathcal{U}_{{\rm R}_y}^{\dag} \tilde{\mathcal {H}}(x,x'; k_y, k_z)  \mathcal{U}_{{\rm R}_y}^{\phantom{\dag}} = \tilde{\mathcal {H}}( x,x' ;-k_y, k_z) ,
\end{align}
 with $\mathcal{U}_{{\rm R}_y} =\delta_{x,x}  \otimes [\mathbb{I}_{2\times 2} \otimes \sigma_2]$, where $ \delta_{x,x}$ acts
on the real-space basis and $\mathbb{I}_{2\times 2} \otimes \sigma_2$ acts on the momentum-space Nambu basis.
Here, $\tilde{\mathcal {H}}(y, y'; k_x, k_z)$ and $\tilde{\mathcal {H}}(x,x'; k_y, k_z)$
 represent the tight-binding Hamiltonians of the $C_{4v}$ NCS in a (010)  and (100) slab geometry, respectively.
Note that the 
arc surface states on the (010) face [(100) face] are left invariant by the
reflection symmetry ${\rm R}_{x}$ [${\rm R}_y$] and that
both reflection operators $\mathcal{U}_{{\rm R}_x}$ and $\mathcal{U}_{{\rm R}_y}$ have eigenvalues $+1$ and $-1$.
 Since the reflection operator $\mathcal{U}_{{\rm R}_x}$ [$\mathcal{U}_{{\rm R}_y}$] commutes with $\tilde{\mathcal{H}} (y,y';k_x=0,k_z)$ [$\tilde{\mathcal{H}} (x,x'; k_y=0,k_z)$],
 the eigenfunctions of  $\tilde{\mathcal{H}} (y,y';k_x=0,k_z)$ [$\tilde{\mathcal{H}} (x,x'; k_y=0,k_z)$] are simultaneous eigenstates of $\mathcal{U}_{{\rm R}_x}$ [$\mathcal{U}_{{\rm R}_y}$]
 with  eigenvalues $+1$ or $-1$. 
 We have checked that the two helical zero-energy surface states
 belong to different eigenspaces of $\mathcal{U}_{{\rm R}_x}$ [$\mathcal{U}_{{\rm R}_y}$]. Thus these zero-energy states cannot hybridize and are protected by these reflection symmetries.


\begin{thebibliography}{999}
\bibitem{Hasan}
M. Z. Hasan and C. L. Kane, 
\newblock  Rev. Mod. Phys. {\bf 82}, 3045 (2010).       

\bibitem{Qi}                     
 X.-L. Qi and S.-C. Zhang, 
\newblock  Rev. Mod. Phys. {\bf 83}, 1057 (2011).    

\bibitem{Ryu2010ten}
S. Ryu, A. P. Schnyder, A. Furusaki, and A. W. W. Ludwig,
\newblock New J. Phys., {\bf 12}, 065010 (2010).

\bibitem{beenakkerReview11}
C. W. J. Beenakker, 
\newblock  Annu. Rev. Condens. Matter Phys. {\bf 2}, 55 (2011).

\bibitem{aliceaReview12}
J. Alicea, 
\newblock  Rep.  Prog.  Phys. {\bf 75}, 6501(2012).

\bibitem{Mourik}                     
V. Mourik, K. Zuo, S. M. Frolov, S. R. Plissard, E. P. A. M. Bakkers, and L. P. Kouwenhoven,
\newblock  Science {\bf 336}, 6084 (2012).   

\bibitem{Das}                     
A. Das, Y. Ronen, Y. Most,	 Y. Oreg, M. Heiblum, and H. Shtrikman,
\newblock  Nat. Phys. {\bf 8}, 887 (2012).  

\bibitem{Deng}                     
M. T. Deng, C. L. Yu, G. Y. Huang, M. Larsson, P. Caroff, and H. Q. Xu,
\newblock  Nano. Lett. {\bf 12}, 6414 (2012).  


\bibitem{Eduardo}                     
E. J. H. Lee, X. Jiang, M. Houzet, R. Aguado, C. M. Lieber, and S. De Franceschi,
\newblock  Nat. Nanotechnol.  {\bf 9}, 79 (2014).  


\bibitem{sau}
J. D. Sau, R. M. Lutchyn, S. Tewari, and S. Das Sarma,
\newblock Phys. Rev. Lett. {\bf 104}, 040502  (2010).
 
\bibitem{lutchyn}
R. M. Lutchyn, J. D. Sau and S. Das Sarma
\newblock Phys. Rev. Lett. {\bf 105}, 077001(2010).
 
\bibitem{oreg} 
Y. Oreg, G. Refael, and F. von Oppen,
\newblock Phys. Rev. Lett. {\bf 105}, 177002 (2010).

\bibitem{fuKanePRL08}
L. Fu  and C. L. Kane, 
\newblock Phys. Rev. Lett. {\bf 100}, 096407 (2008).

\bibitem{bauerSigristBook}
E. Bauer and M. Sigrist, 
\newblock
\textit{Non-Centrosymmetric Superconductors: Introduction and Overview},
Lect. Notes Phys. {\bf 847}, 1-357 (2012).

\bibitem{SatoPRB06}
M. Sato, 
\newblock Phys. Rev. B {\bf 73} 214502 (2006).

\bibitem{Tanaka}
Y. Tanaka, Y. Mizuno, T. Yokoyama, K. Yada, and M. Sato,
\newblock Phys. Rev. Lett. {\bf 105} 097002 (2010).

\bibitem{Sato2} 
M. Sato, and S. Fujimoto,
\newblock Phys. Rev. Lett. {\bf 105} 217001 (2010).

\bibitem{Beri}
B. B\'eri,
\newblock  Phys. Rev. B {\bf 81} 134515 (2010).

\bibitem{Schnyder}
A. P. Schnyder and S. Ryu, 
\newblock Phys. Rev. B {\bf 84} 060504 (2011).

\bibitem{Yada}
K. Yada, M. Sato, Y. Tanaka, and T. Yokoyama,
\newblock  Phys. Rev. B {\bf 83} 064505 (2011).

\bibitem{Brydon}
P. M. R. Brydon, A. P. Schnyder, and C. Timm,  
\newblock Phys. Rev. B {\bf 84} 020501 (2011).

\bibitem{Matsuura}
S. Matsuura, P. -Y. Chang, A. P. Schnyder, S. Ryu,
\newblock  New J. Phys. {\bf 15} 065001 (2013).

\bibitem{Schnyder1}
A. P. Schnyder, P. M. R. Brydon, and C. Timm,
\newblock  Phys. Rev. B {\bf 85} 024522 (2012).

 \bibitem{Sato3} 
M. Sato and S. Fujimoto,
\newblock Phys. Rev. B {\bf 79} 094504 (2009).

\bibitem{bauer2004heavy}
E.\ Bauer, G. Hilscher, H. Michor, Ch. Paul, E. W. Scheidt, A. Gribanov, Yu. Seropegin, H. Noe?l, M. Sigrist, and P. Rogl,
\newblock Phys. Rev. Lett. {\bf 92} 027003 (2004).

\bibitem{Izawa}
K. Izawa, Y. Kasahara, Y. Matsuda, K. Behnia, T. Yasuda, R. Settai, and Y. Onuki, 
\newblock   Phys. Rev. Lett. {\bf 94}, 197002 (2005).

\bibitem{Bonalde}
I. Bonalde, R. L. Robeiro, W. Br\"amer-Escamilla, C. Rojas, E. Bauer, A. Prokofiev, Y. Haga, T. Yasuda, and Y. \={O}nuki, 
\newblock  New. J. Phys. {\bf 11}, 055054 (2009).

\bibitem{Mukuda}  
H. Mukuda, T. Fujii, T. Ohara, A. Harada, M. Yashima, Y. Kitaoka, Y. Okuda, R. Settai, and Y. Onuki, 
\newblock Phys. Rev. Lett. {\bf 100}, 107003 (2008).

\bibitem{kimuraPRL05}
N. Kimura, K. Ito, K. Saitoh, Y. Umeda, H. Aoki, and T. Terashima, 
\newblock Phys. Rev. Lett. {\bf 95}, 247004 (2005).

\bibitem{chenPRB11}
J.  Chen, M. B. Salamon, S. Akutagawa, J. Akimitsu, J. Singleton, J. L. Zhang, L. Jiao, and H. Q. Yuan, 
\newblock Phys. Rev. B {\bf 83}, 144529 (2011).

\bibitem{Yuan}  
 H. Q. Yuan, D. F. Agterberg, N. Hayashi, P. Badica, D. Vandervelde, K. Togano, M. Sigrist, and M. B. Salamon, 
\newblock Phys. Rev. Lett. {\bf 97}, 017006 (2006).

\bibitem{Nishiyama}  
M. Nishiyama, Y. Inada, and G.-Q. Zheng, 
\newblock Phys. Rev. Lett. {\bf 98}, 047002 (2007).

\bibitem{Eguchi}  
G. Eguchi, D. C. Peets, M. Kriener, S. Yonezawa, G. Bao, S.
Harada, Y. Inada, G.-q. Zheng, and Y. Maeno, 
\newblock Phys. Rev. B {\bf 87}, 161203(R) (2013).

\bibitem{Mondal}
M. Mondal, B. Joshi, S. Kumar, A. Kamlapure, S. C. Ganguli, A. Thamizhavel, S. S. Mandal, S. Ramakrishnan, and P. Raychaudhuri, 
\newblock Phys. Rev. B {\bf 86}, 094520 (2012).

\bibitem{Sato1}
M. Sato, Y. Tanaka, K. Yada, and T. Yokoyama,
\newblock Phys. Rev. B {\bf 83} 224511 (2011).

\bibitem{HofmannPRB13}
J. S. Hofmann, R. Queiroz, and A. P. Schnyder,
\newblock  Phys. Rev. B {\bf 88}, 134505 (2013).

\bibitem{brydonNJP13}
P. M. R. Brydon, C. Timm, A. P. Schnyder,
\newblock New J. Phys. {\bf 15}, 045019 (2013).

\bibitem{Schnyder2}
A. P. Schnyder, Carsten Timm, and P. M. R. Brydon,
\newblock Phys. Rev. Lett. {\bf 111} 077001 (2013).

\bibitem{Lu}  
C. -K. Lu and S. Yip,
 \newblock Phys. Rev. B {\bf 78} 132502 (2008).

\bibitem{Fuji08}  
S. Fujimoto, 
\newblock Phys. Rev. B {\bf 77} 220501(R) (2008).

\bibitem{Kashyap}
M. K. Kashyap and D. F. Agterberg, 
\newblock Phys. Rev. B {\bf 88} 104515 (2013).

\bibitem{footnote_vortex}
For previous studies on stable Majorana vortex-bound states 
in fully gapped 2D NCSs, 
in 3D SCs with Rashba-spin orbit coupling,
and NCSs with cubic crystal symmetry $O$,
see Refs.~\onlinecite{Fuji08, Sato3, Lu, Kashyap}.

\bibitem{footnote_T_breaking}
This is in some sense expected since 
helical modes are unstable against TRS breaking perturbation,
and here vortices break TRS.  
  

 
  
\bibitem{Wan}
X. Wan, A. M. Turner, A. Vishwanath, and S. Y. Savrasov,
\newblock Phys. Rev. B {\bf 83}, 205101 (2011).

\bibitem{Burkov}
A. A. Burkov and Leon Balents,
\newblock Phys. Rev. Lett. {\bf 107}, 127205 (2011).

\bibitem{Xu}
G. Xu, H. Weng, Z. Wang, X. Dai, and Z. Fang, 
\newblock Phys. Rev. Lett. {\bf 107}, 186806 (2011).







  
\bibitem{frigeri04}
P. A. Frigeri, D. F. Agterberg, A. Koga, and M. Sigrist, Phys. Rev.
Lett. {\bf 92}, 097001 (2004).

\bibitem{Samokhin}
K.V. Samokhin,
\newblock Annals of Physics {\bf 324} 2385 (2009).
   

\bibitem{gapclosing}
The gap closes when $0=|B-A  | {\bf l}  |  |$ with $B=\varepsilon({\bf k} )+i \Delta_s$ and $A=\alpha+i \Delta_t$. 
The gapless points are the intersections between hypersurfaces that are characterized by these two equations:
$t(\cos k_x+\cos k_y+\cos k_z)=\mu+\alpha \Delta_s/\Delta_t$ and 
$| {\bf l}  |=\Delta_s/\Delta_t$.   





\bibitem{continuum}
Following Ref.~\onlinecite{Lu},
we consider the normal state band structure 
$h = (-\frac{\nabla^2}{2m}-\mu) \mathbb{I}_{2 \times 2}+\alpha {\bf l}( {\bf k} ) \cdot \pmb{\sigma}$
and the pairing term 
$\Delta=(\Delta_s+\frac{1}{2} \pmb{\nabla}\cdot {\bf D} + {\bf D} \cdot \pmb{\nabla}) (i \sigma_2)$,
where $m$ is the effective mass and ${\bf D}= - i \pmb{\nabla}_{\bf k} (\Delta_p {\bf l}( {\bf k} ) \cdot \pmb{\sigma})$.
We can introduce a vortex line along the $z$ axis localized at the origin by adding a phase on gap functions for both singlet and triplet pairings.
In the continuum model, we need to linearize ${\bf l}( {\bf k} )= (a_1 k_x, a_1 k_y, a_3 k_z)$.
In addition, we consider the asymptotic limit ($1/r \to 0$) that we can neglect all $\frac{1}{r}$ and $\frac{1}{r^2}$ terms in the continuum BdG equation,
where $r$ is the radial direction in the cylindrical coordinate.
As the result, there is a localized zero energy solution that decays as a function of $r$.

\bibitem{FangPRL14}
C. Fang, M. J. Gilbert, and B. A. Bernevig, Phys. Rev. Lett. {\bf 112}, 106401 (2014).

\bibitem{MizushimaPRL12}
T. Mizushima, M. Sato, and K. Machida, Phys. Rev. Lett. {\bf 109}, 165301 (2012).

\bibitem{KenArXiv} 
K. Shiozaki and M. Sato, Phys. Rev. B {\bf 90}, 165114 (2014).  




\bibitem{footnote_Weyl}  
The Fermi arcs discussed in Refs. \onlinecite{Wan, Burkov, Xu}
are protected by a $\mathbb{Z}$ number. In our case, due to time-reversal
symmetry, the helical arc surface states are protected by a $\mathbb{Z}_2$ number.

\bibitem{Ojanen}
T. Ojanen,
\newblock Phys. Rev. B {\bf 87}, 245112 (2013).


\bibitem{Morimoto}
T. Morimoto and A. Furusaki,
\newblock Phys. Rev. B {\bf 89}, 235127 (2014).

\bibitem{volovik11}
G. E. Volovik,
\newblock  JETP Lett. {\bf 93} 66 (2011).

\bibitem{volovikBOOKS}
G. E. Volovik, in 
\textit{The Universe in a Helium Droplet}, 
The International Series of Monographs on Physics Vol. 117 (Oxford University Press, New York, 2003); 
G. E. Volovik, in 
\textit{Exotic Properties of Superfluid 3He}, Series I N Modern Condensed Matter Physics Vol. 1 
(World Scientific, Singapore, 1992).

\bibitem{footnote_bdg_cont2}
For the $C_2$ point-group NCS with $a_4 \ne 0$ and $a_5 \ne 0$, the variables $r$ and $\theta$ are not  
separable in the continuum BdG equation given in the  
Appendix~\ref{App: Continuous model}, which  may imply
that there are no zero-energy solutions localized at the vortex core. 


\bibitem{Slager1}
R.-J. Slager, A. Mesaros, V. Juricic, and J. Zaanen, arXiv:1401.4044.

\bibitem{Sasaki}
S. Sasaki, K. Segawa, and Y. Ando,
arXiv:1404.1707.

\bibitem{zhangeKaneMele}
F. Zhang, C. L. Kane, and E. J. Mele, Phys. Rev. Lett. {\bf 111}, 056402 (2013).



\bibitem{SunArXiv14}
Z.\ Sun, M. Enayat, A. Maldonado, C. Lithgow, E. Yelland, D. C. Peets, A. Yaresko, A. P. Schnyder, and P. Wahl, arXiv:1407.5667.

\bibitem{Nomura}
K. Nomura, S. Ryu, A. Furusaki, and N. Nagaosa,
\newblock Phys. Rev. Lett. {\bf 108}, 026802 (2012).

\end{thebibliography}
\end{document}